\documentclass[aps,prb,twocolumn,showpacs,amsmath]{revtex4}
 \usepackage{graphics}
 \usepackage{epsfig}
\begin{document} 
   \title{\Large \bf 
Dynamic scattering channels of a double barrier structure }
\author{M. Moskalets$^{1,2}$ and M. B\"uttiker$^1$}
\affiliation{
$^1$D\'epartement de Physique Th\'eorique, Universit\'e de Gen\`eve, CH-1211 Gen\`eve 4, Switzerland\\
$^2$Department of Metal and Semiconductor Physics,\\
National Technical University "Kharkiv Polytechnic Institute",
            61002 Kharkiv, Ukraine\\
}
\date\today
   \begin{abstract}
We calculate analytically the Floquet scattering matrix for a periodically driven double-barrier structure.
Our approach takes into account dynamical effects which become necessarily important when electrons propagate through a system subject to a fast drive. 
It is convenient to represent the Floquet scattering amplitude as a sum of amplitudes corresponding to different times spent by an electron inside the structure. 
These amplitudes define dynamical scattering channels.
Then we represent the dc current generated by the structure as a sum of two generic contributions.
The first one is due to photon-assisted interference processes within the same channel
and the second one is due to inter-channel interference processes.
At zero temperature both contributions are present
while at high temperatures and/or high driving frequencies only the former survives.
\end{abstract}
\maketitle
\small

\section{Introduction}
\label{intro}
Conducting structures subject to periodically varying voltages or magnetic fields are of a fundamental interest. A phase-coherent conductor subject to a slow ac drive \cite{BTP94} under quite general conditions \cite{Brouwer98} becomes a quantum pump 
\cite{AA98}$^-$ \cite{CC08}
which is able to generate a dc current. 
Such a current generated by an unbiased mesoscopic conductor was detected experimentally. 
\cite{SMCG99}$^-$\cite{VDM05}
Theoretically the quantum pumps are analyzed for both slow \cite{RB06}$^-$\cite{SGK08}
and high 
\cite{TC01}$^-$\cite{BB08}
driving frequencies. 

Compared to adiabatic pumps, high-frequency pumps have the advantage that the currents which are generated are much larger. In addition, at low frequencies a proper pump current is difficult to distinguish from a current generated by rectification \cite{DMH03,VDM05}. 
In contrast high-frequency pumps have been realized in recent experiments \cite{2bpumpexp07}$^-$\cite{KKHPSS08}.  These pumps 
might find an application for metrological purposes \cite{PG00} and or as an on demand electron source for solid state systems for future quantum information processing.
Possibly another application is the generation of correlated particles \cite{SamuelssonB04}$^-$\cite{MBen05} of a pump outside the quantized pumping regime.

Though several analytical results are available \cite{LN93}$^-$\cite{DO08},  
the theoretical analysis of  pumps at high driving frequencies is based mainly on numerical calculations \cite{CTCC04}$^-$\cite{AYM-R07}. 
The aim of the present paper is to develop an analytical approach allowing to describe the scattering by mesoscopic systems at arbitrary driving frequencies. 
We consider a model consisting of two barriers with varying strength and with a uniform varying potential in between them.
Such a structure can serve as a simple model of an actual structure used in experiments \cite{2bpumpexp07}$^-$\cite{KKHPSS08} and can serve as a prototype of a quantum pump with immovable in space potentials. 
Our approach is based on the Floquet scattering matrix formalism which we developed in Ref.\,\onlinecite{MBstrong02}

\begin{figure}[t]
  \vspace{0mm}
  \centerline{
   \epsfxsize 8cm
   \epsffile{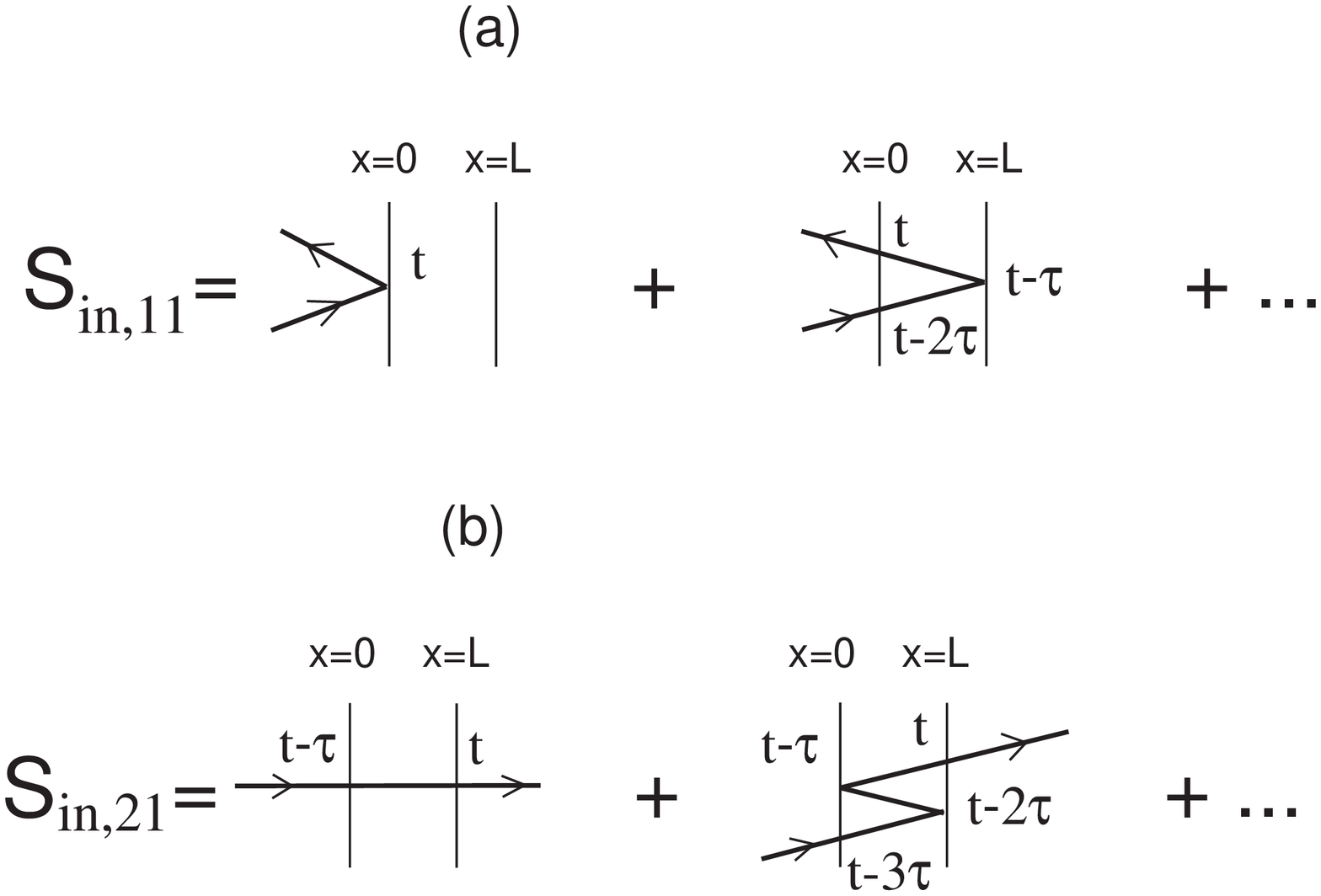}
             }
  \vspace{0mm}
  \nopagebreak
  \caption{Some processes contributing to reflection amplitude $S_{in,11}$ (a) and to transmission amplitude $S_{in,21}$ (b) are depicted. The point-like scatterers located at $x=0$ and $x=d$ are shown by vertical lines. The particle leaves the system at time moment $t$. We also indicate the time moments when the reflection/transmission at point-like scatterers occurs. $\tau$ is the propagation time from one point-like scatterer to the other.}
\label{fig1}
\end{figure}

The paper is organized as follows.
In Sec.\ref{DB} we represent the Floquet scattering matrix of a fast driven double-barrier system  as a sum of the contributions arising from different reflection/transmission processes accompanying an electron propagation through the system, Fig.\,\ref{fig1}. 
Our approach takes into account effects due to finiteness of the time spent by electrons inside the system. In Sec.\ref{dc}
we find the dc generated current at arbitrary driving frequency and arbitrary temperature.
The correlation properties of currents which are generated by the pump are explored in Sec.\ref{zfn}.
The currents flowing through the structure which in addition is subject to an external dc or ac bias are calculated in Sec.\ref{bias}. 
The discussion of our results and a conclusion are given in Sec.\ref{concl}.

\section{Floquet scattering matrix}
\label{DB}
It is convenient to introduce the matrix $\hat S_{in}(t,E)$ which gives the current amplitude of particles exciting the scatterer at time t if they have been incident (index in) on the scatterer with energy $E$.  The Fourier coefficients of this matrix define the Floquet scattering matrix $\hat S_{F}$: 

\begin{equation}
\label{Eq_new1}
	\hat S_{F}(E_{n},E) = \int\limits_{0}^{\cal T}
	\frac{dt}{\cal T} e^{in\Omega t} \hat S_{in}(t,E)\,.
\end{equation}
\ \\
Here ${\cal T} = 2\pi/\Omega$ is the period of a parameter which describes the modulation of the scattering matrix, and 
$E_n = E + n\hbar\Omega$, is the energy of carriers which have absorbed $n$ modulation quanta. 
$n$ is an integer.

For $\hbar\Omega\ll E$ one can obtain an analytical expression for $\hat S_{in}(t,E)$ of a periodically driven system consisting of two point-like scatterers placed at a distance $d$ of each other with a potential $U$ in between them. For a structure with contacts labeled $\alpha, \beta,..$ 
the matrix element $S_{in,\alpha\beta}(t,E)$ can be given as a sum over different paths ${\cal L}_{\alpha\beta}^{(q)}$ which an electron incident in contact $\beta$ can follow to propagate through the system leaving it at contact $\alpha$ with either transmission or reflection at any point-like scatterer. 
Thus ${\cal L}_{\alpha\beta}^{(q)}$ represents a classification of paths according to contacts and the number of reflections. 
We introduce a partial amplitude ${\cal S}_{\alpha\beta}^{(q,\tau)}(t,E)$ which describes a process during which a particle with energy $E$ enters the system through the lead $\beta$, undergoes $2q-\delta_{\alpha\beta}$ (for $q\ne 0$) reflections, and leaves the system through the lead $\alpha$ at a time moment $t$, see Fig.\,\ref{fig1}. Carriers take a time $\tau = d/v$ to propagate from one barrier to the other. 
This amplitude is a product of amplitudes corresponding to an instantaneous interaction (reflection/transmission) with time-dependent point-like barriers and amplitudes 

\begin{equation}
e^{i[kd\, -\, (e/\hbar)\,\int^{t_j}_{t_j-\tau}\, dt^\prime\, U(t^{\prime})]} 
\label{Eqmb1}
\end{equation}
\ \\
corresponding to a free propagation 
(starting at time $t_j-\tau$ and  having a duration of $\tau=d/v$) between the two barriers. 
The time moments, when the instantaneous reflection/transmission amplitudes are calculated, are counted backward along the path in a descending order starting from the time moment $t$ when the particle leaves the system. The detailed calculation is presented in 
Appendix~\ref{appendix0}. We obtain 

\begin{equation}
\label{Eqdb_13A}
S_{in,\alpha\beta}(t,E) = \sum\limits_{q=0}^{\infty}\, e^{2i\,q_{\alpha\beta}\,kd}\,  {\cal S}_{\alpha\beta}^{(q,\tau)}(t,E)\,, 
\end{equation}
\ \\
where $2q_{\alpha\beta} = 2q+1-\delta_{\alpha\beta}$,
$k$ is the wave number, and the quantities ${\cal S}_{\alpha\beta}^{(q,\tau)}(t,E)$ are given in Eq.\,(\ref{Eqdb_13B}).

The matrix $\hat S_{in}(t)$ is periodic in time. This is due to the time-periodicity of the frozen scattering matrices $\hat L$, $\hat R$ which describe transmission and reflection at the left and right barriers. Here we take these barriers to be point-like scatterers. 
In addition  $\hat S_{in}$ is periodic in the driving frequency $\Omega$.
This follows from the observation that Eqs.\,(\ref{Eqdb_13A}), (\ref{Eqdb_13}) give the same $\hat S_{in}$ for $\tau$ being a multiple of a driving period: $\tau = n{\cal T}$, where $n$ is an integer. Therefore, $\hat S_{in}$ is the same for any $\Omega = 2\pi n\tau^{-1}$.

The Floquet scattering matrix is unitary:

\begin{subequations}
\label{Eq_new2}
\begin{equation}
\label{Eq_new2A}
\sum\limits_{n=-\infty}^{\infty} {\hat S}_{F}^{\dagger}(E_{m},E_{n})
{\hat S}_{F}(E_{n},E) = \hat I\delta_{m,0}\,,
\end{equation}

\begin{equation}
\label{Eq_new2B}
\sum\limits_{n=-\infty}^{\infty} {\hat S}_{F}(E_{m},E_{n})
{\hat S}_{F}^{\dagger}(E_{n},E) = \hat I\delta_{m,0}\,.
\end{equation}
\end{subequations}
\ \\
Here $\hat I$ is a unit matrix. 
Substituting Eq.\,(\ref{Eq_new1}) into Eq.\,(\ref{Eq_new2A}) we get the following unitarity condition for the matrix $\hat S_{in}(t,E)$:\cite{MBmag05} 

\begin{equation}
\label{Eqdb_15}
\int\limits_{0}^{\cal T} \frac{dt}{{\cal T}} 
\hat S_{in}^{\dagger}(E,t) \hat S_{in}(t,E) = \hat I\,.
\end{equation}

Note in Ref.\,\onlinecite{MBmag05} only the low frequency, $\Omega\to 0$, asymptotics for $\hat S_{in}(t,E)$ was calculated. 
In contrast equation (\ref{Eqdb_13A}) is valid at any frequency $\Omega\ll\mu$ of a drive.
In the Appendix~\ref{appendix} we make a connections between these two calculations.

\section{dc generated current}
\label{dc}

If the Floquet scattering matrix $\hat S_{F}$ is known then the dc current $I_{\alpha}$ flowing out of the structure through the lead $\alpha$ can be calculated. To be definite we consider the usual set-up for a mesoscopic electron quantum pump. The pump connects $N_r$ reservoirs. 
The reservoirs are in stationary equilibrium states and have the same chemical potentials, $\mu_{\alpha}=\mu$, and temperatures, $T_{\alpha}=T$. 
Then the distribution function $f_{\alpha}$ of reservoir $\alpha$ is the Fermi distribution function $f_{\alpha}(E) = f_{0}(E) \equiv\left[1+ \exp\left(\frac{E - \mu}{k_BT}\right)\right]^{-1}$, where $k_B$ is the Boltzmann constant. The current is \cite{MBstrong02,MBac04}

\begin{equation}
\label{Eqdc_1}
\!\!I_{\alpha}\! =\! \frac{e}{h}\int\!\! dE \!\sum\limits_{\beta=1}^{N_r}
\sum\limits_{n=-\infty}^{\infty}
\left|S_{F,\alpha\beta}(E_n, E)\right|^2
\left[f_{\beta}(E) - f_{\alpha}(E_n) \right]\,.
\end{equation}
\ \\
Using Eqs.\,(\ref{Eq_new1}), (\ref{Eqdb_13A}) and (\ref{Eqdb_13}) one can calculate the dc current generated by the scatterer under consideration. 

To proceed analytically we assume the temperature to be smaller than the Fermi energy,

\begin{equation}
\label{Eqdc_2}
k_BT \ll \mu\,.
\end{equation}
\ \\
Then calculating Eq.\,(\ref{Eqdc_1}) we can safely neglect the dependence on energy of the scattering matrices of the point-like scatterers, see Eqs.\,(\ref{Eqdb_13}), and evaluate them at the Fermi energy, $E=\mu$. 
In addition we approximate the phase factor with a linear in energy expansion away from its value at the Fermi energy

\begin{equation}
\label{Eqdc_3}
e^{iqkd} \approx e^{iqk_Fd}e^{iq\frac{E-\mu}{\hbar}\tau_0}\,.
\end{equation}
\ \\
Thus from now on the propagation time is taken $\tau = \tau_0 = d/v_F$, with $v_F$ the velocity of an electron with the Fermi energy.

With these approximations we perform an energy integration in Eq.\,(\ref{Eqdc_1}) and find the dc current as a sum of a contribution $I_{\alpha}^{(0)}$ diagonal in dynamical scattering channels and a contribution $I_{\alpha}^{(i)}$ arising from interference of dynamical scattering channels,  

\begin{subequations}
\label{Eqdc_4}
\begin{equation}
\label{Eqdc_4A}
I_{\alpha} = I_{\alpha}^{(0)} + I_{\alpha}^{(i)}\,. 
\end{equation}
\ \\
The diagonal part consists of contributions of different scattering channels that can be labeled by an integer number $q$ which is related to the number of times a carrier is reflected inside the structure,

\begin{equation}
\label{Eqdc_4B}
 I_{\alpha}^{(0)} = \sum\limits_{q=0}^{\infty}
 J_{\alpha}^{(q)}\,,
\end{equation}
\ \\
with the contribution of the $q-th$ scattering channel given by

\begin{equation}
\label{Eqdc_4C}
 J_{\alpha}^{(q)} = - i\frac{e}{2\pi} \int\limits_{0}^{\cal T} \frac{dt}{\cal T} 
\left( \hat {\cal S}^{(q,\tau_0)}(t,\mu) \frac{\partial \hat{\cal S}^{(q,\tau_0)\dagger}(t,\mu)}{\partial t}\right)_{\alpha\alpha}\,.
\end{equation}
\ \\
The interference contribution is a sum of thermally smeared non-diagonal currents 

\begin{equation}
\label{Eqdc_4D}
I_{\alpha}^{(i)} = {\rm Re}
\sum\limits_{q=1}^{\infty}\sum\limits_{q^\prime=0}^{q-1}
e^{2i(q-q^\prime)k_Fd}\eta(q-q^\prime)J_{\alpha}^{(q,q^\prime)}\,, 
\end{equation}
\ \\
with
\begin{equation}
\label{Eqdc_4E}
\begin{array}{c} 
J_{\alpha}^{(q,q^\prime)}= -i\frac{e}{2\pi}
\int\limits_{0}^{\cal T} \frac{dt}{\cal T} \\
\ \\
\!\!\!\!\times
\left( \hat {\cal S}^{(q,\tau_0)}(t,\mu) \frac{\hat {\cal S}^{(q^\prime,\tau_0)\dagger}(t,\mu)  - \hat {\cal S}^{(q^\prime,\tau_0)\dagger}(t-2\tau_0[q-q^\prime],\mu)}{\tau_0(q-q^\prime)}\right)_{\alpha\alpha}\,, \\
\ \\
\eta(q-q^\prime) = \frac{2\pi k_BT(q-q^\prime)}{\hbar\tau_0^{-1}}
\sinh^{-1}\left( \frac{2\pi k_BT}{\hbar\tau_0^{-1}}[q-q^\prime] \right)\,.
\end{array}
\end{equation}
\end{subequations}
\ \\
The factor $\eta$ describes the effect of averaging over different energies of incoming electrons within a relevant energy window near the Fermi energy. 
Note from Eq.(\ref{m8_3_A}) it follows that the quantity $I_{\alpha}^{(0)}$, Eq.\,(\ref{Eqdc_4B}), is real.

The time of flight $\tau_0 = d/v_F$ plays a twofold role in the problem under consideration. 
On one hand, this time separates adiabatic ${\cal T} \gg \tau_0$ and non-adiabatic ${\cal T} \le \tau_0$ regimes. 
On the other hand it defines the energy scale $T^{\star} = \Delta/(2\pi^2k_B)$ separating  low and high temperature regimes. 
Here we have introduced the level spacing $\Delta = \pi\hbar\tau_0^{-1}$ (near the Fermi energy) for a double-barrier structure detached from the leads. 
At low temperatures, $T \ll T^{\star}$, we have $\eta = 1$, 
while at high temperatures, $T \gg T^{\star}$, the factor $\eta$ is exponentially small, 

\begin{equation}
\label{new_01}
\eta \approx 2|q-q\prime|\frac{T}{T^\star}e^{-\frac{T}{T^\star}|q-q\prime|} 
\end{equation}
\ \\
and describes the exponential suppression of the interference terms at large temperatures.  

Note the effect of the temperature we consider is only due to energy averaging. We do not consider dephasing but assume that electrons preserve phase coherence as they propagate through the sample. 
The crossover temperature $T^{\star}$ is familiar from the persistent current problem (see, e.g., Ref.\,\onlinecite{CGRH88}) and it appears as a characteristic temperature \cite{MoskaletsIBS98} for interference phenomena in single channel ballistic structures.

The two parts, $I_{\alpha}^{(0)}$ and $I_{\alpha}^{(i)}$, of the pumped current result from different processes that lead to different temperature dependencies. \cite{note1}
The first part, $I_{\alpha}^{(0)}$, is a sum of contribution, $J_{\alpha\beta}^{(q)}$, arising from different electron's paths inside the system. The paths differ by incoming ($\beta$) and outgoing ($\alpha$) leads, and by the index $q$ counting the number of reflections inside the system.
According to Ref.\,\onlinecite{BM05} the quantum pump effect (i.e., the effect which results in a dc current generated by the periodically driven mesoscopic system) is due to interference of various photon-assisted amplitudes describing the interaction of propagating electrons with a scatterer that varies periodically in time. 
Therefore, one can consider the contribution $J_{\alpha\beta}^{(q)}$ as due to photon-assisted interference processes taking place at the same spatial path ${\cal L}_{\alpha\beta}^{(q)}$. 
Each such path can be characterized by a delay time $2q_{\alpha\beta}\tau$, i.e., the difference of times when an electron enters and leaves the system. 
If this time is not small compared with the driving period then dynamical effects become important for an electron scattering off the system.
Therefore, one can consider the path ${\cal L}_{\alpha\beta}^{(q)}$ as an effective {\it dynamical scattering channel}. 
Then we interpret $J_{\alpha}^{(q)}$ as arising due to intra-channel photon-assisted interference processes. 
Since all the quantum-mechanical amplitudes corresponding to such processes are multiplied by same dynamical factor 
$e^{2iq_{\alpha\beta}kd}$,
the corresponding probability is independent of energy.
Consequently the energy integration becomes trivial.

In contrast, the second part, $I_{\alpha}^{(i)}$, due to interference between  different paths (i.e., due to inter-channel interference) is defined as a sum of terms oscillating in energy.
Consequently it vanishes at high temperatures.

Equation (\ref{Eqdc_4})  gives the dc current generated by the double-barrier structure at arbitrary temperature and at arbitrary driving frequency.
At slow driving, ${\cal T} \gg \tau$, it agrees with the famous Brouwer's result \cite{Brouwer98} for an adiabatic pumped current. 
At larger driving frequencies, it reproduces the numerical results of Ref.\,\onlinecite{MBstrong02} obtained for a double-barrier structure in the zero temperature limit. 

\begin{figure}[t]
  \vspace{0mm}
  \centerline{
   \epsfxsize 8cm
   \epsffile{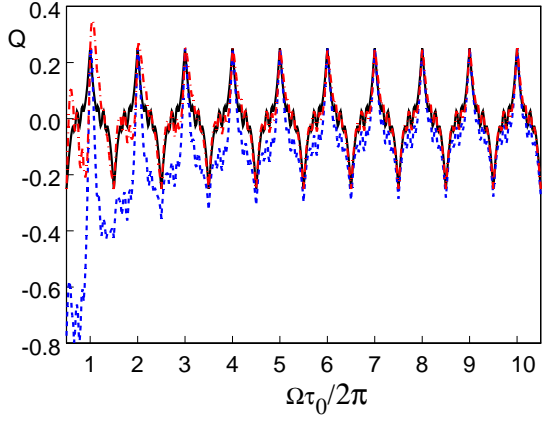}
             }
  \vspace{0mm}
  \nopagebreak
  \caption{(color online) The charge $Q$ = $2\pi I^{(0)}/\Omega$ in units of $e$ pumped for a cycle as a function of $\Omega$. The black solid line is for $eU=0$. The red dash-dotted (blue dashed) line is for $eU=5\Delta$ and $\varphi_U = \pi/4$ ($\varphi_U = 0$). The parameters are: 
$L = 200\pi$; 
$V_{01} = V_{02} = 20$; 
$V_{11} = V_{12} = 10$; 
$\varphi_1=0$;
$\varphi_2=\pi/2$.
We use the units: $2m = \hbar = e = 1$.
The Fermi energy is $\mu = 17.4\,$.}
\label{fig2}
\end{figure}

From Eq.\,(\ref{Eqdc_4}) it follows that with increasing temperature or driving frequency the different dynamical scattering channels contribute independently to the dc current, $I_{\alpha}\approx I_{\alpha}^{(0)}$. 
With regard to the frequency this follows from the observation that the ratio $I_{\alpha}^{(0)}/I_{\alpha}^{(i)}$ behaves as $\Omega\tau_0$.
Therefore, at $\Omega\to\infty$ the contribution $I_{\alpha}^{(0)}$ dominates. 
So we can conclude: 
(i) At high temperatures, $T\gg T^\star$ the dc current (both adiabatic and non-adiabatic) is independent of the Fermi energy (over a scale much smaller than the Fermi energy itself). 
(ii) At high driving frequencies, $\Omega\gg \pi\tau_0^{-1}$, the dc current becomes independent of the temperature.
The latter is illustrated in Fig.\,\ref{fig2} (compare with Fig.\,3 in Ref.\,\onlinecite{MBstrong02}), where we give the high-temperature dc current $I_{\alpha}^{(0)}$ for a double-barrier structure with and without the uniform oscillating potential $U(t)$ as a function of the driving frequency $\Omega$.
One can see, that the presence
of an oscillating uniform potential does not change the universal high-frequency behavior of the resulting dc current. 

\section{Zero frequency noise power}
\label{zfn}

We consider the zero-frequency Fourier coefficient of the correlation function of currents generated by the dynamic scatterer:

\begin{equation}
\label{Eqzfn_1}
{\bf P}_{\alpha\beta} = \int\limits_{0}^{\cal T} \frac{dt}{\cal T} 
\int\limits_{-\infty}^{\infty} dt^{\prime} 
\langle
\Delta\hat I_{\alpha}(t)\Delta\hat I_{\beta}(t^{\prime}) +
\Delta\hat I_{\beta}(t^{\prime})\Delta\hat I_{\alpha}(t) \rangle\,,
\end{equation}
\ \\
where 
$\Delta\hat I = \hat I - \langle\hat I\rangle$, and $\langle\cdots\rangle$ denotes quantum-statistical averaging.
This quantity can be represented as a sum of a thermal (Nyquist-Johnson) noise ${\bf P}_{\alpha\beta}^{(th)}$ and a shot noise ${\bf P}_{\alpha\beta}^{(sh)}$, ${\bf P}_{\alpha\beta} = {\bf P}_{\alpha\beta}^{(th)} + {\bf P}_{\alpha\beta}^{(sh)}$.
Notice, each of these contributions depends on both the temperature and the drive. 
The reason to divide the whole noise into two parts is rather conventional: 
The former contribution vanishes at zero temperature while the latter one vanished in the absence of a drive. 
In terms of the elements the matrix $\hat S_{in}(t,E)$ the noise is \cite{MBnoise04}

\begin{subequations}
\label{Eqzfn_2}
\begin{equation}
\label{Eqzfn_2A}
\begin{array}{c}
{\bf P}_{\alpha\beta}^{(th)} = \frac{2e^2k_BT}{h}\int dE \left( - \frac{\partial f_0}{\partial E}\right)
\int\limits_{0}^{\cal T} \frac{dt}{\cal T} 
\bigg(- \left|S_{in,\alpha\beta}(t,E) \right|^2 \\
\ \\ 
- \left|S_{in,\beta\alpha}(t,E) \right|^2 
+ \delta_{\alpha\beta} \Big[1 + {\it \Sigma}_{in,\alpha\alpha}(t,t;E)
\Big]
\bigg)\,,
\end{array}
\end{equation}

\begin{equation}
\label{Eqzfn_2B}
\begin{array}{c}
{\bf P}_{\alpha\beta}^{(sh)} = \frac{e^2}{h}\int dE
\sum\limits_{m=-\infty}^{\infty} 
\left[f_0(E) - f_0(E_m) \right]^2 
\int\limits\!\!\! \int\limits_{0}^{\cal T} \frac{dt_1 dt_2}{{\cal T}^2} 
\\
\ \\
\times
e^{im(t_2 - t_1)}
{\it \Sigma}_{in,\alpha\beta}^{\star}(t_1,t_2;E)
{\it \Sigma}_{in,\alpha\beta}(t_1,t_2;E_m)\,.
\end{array}
\end{equation}
\end{subequations}
\ \\
Here $\hat {\it \Sigma}_{in}$ is the two-particle scattering matrix: \cite{MBnoise04}

\begin{equation}
\label{Eqzfn_3}
\hat {\it \Sigma}_{in}(t_1,t_2;E) = \hat S_{in}(t_1,E)\hat S_{in}^{\dagger}(E,t_2)\,.
\end{equation}

To calculate the noise power generated by a driven double-barrier structure we use Eqs.\,(\ref{Eqdb_13A}), (\ref{Eqdc_2}), (\ref{Eqdc_3}) and (\ref{Eqdb_13}).

\subsection{Thermal noise}

The thermal noise can be further separated into two contributions  
${\bf P}_{\alpha\beta}^{(th)} = {\bf P}_{\alpha\beta}^{(th,0)} + {\bf P}_{\alpha\beta}^{(th,i)}$, 
where 

\begin{subequations}
\label{Eqzfn_4}
\begin{equation}
\label{Eqzfn_4A}
\begin{array}{c}
{\bf P}_{\alpha\beta}^{(th,0)} = \frac{2e^2}{h}k_BT \left( 2\delta_{\alpha\beta} - g_{\alpha\beta} -  g_{\beta\alpha} \right)\,, \\
\ \\
g_{\alpha\beta} = 
\int\limits_{0}^{\cal T} \frac{dt}{\cal T} g_{\alpha\beta}(t)\,,  
\quad
g_{\alpha\beta}(t) = 
\sum\limits_{q=0}^{\infty}
\left|{\cal S}_{\alpha\beta}^{(q,\tau_{0})}(t,\mu) \right|^2
\end{array}
\end{equation}
\ \\
is the contribution that survives at high temperatures. 
Note that the (dimensionless) matrix $\hat g$ and hence ${\bf P}^{(th,0)}$ is independent of the oscillating potential $U(t)$ of the well.

At low temperatures, $T\ll T^\star$ there exists an additional contribution ${\bf P}_{\alpha\beta}^{(th,i)}$ to thermal noise:

\begin{equation}
\label{Eqzfn_4B}
\begin{array}{c}
{\bf P}_{\alpha\beta}^{(th,i)} = \frac{4e^2}{h} k_BT\, {\rm Re} \sum\limits_{q=1}^{\infty}\sum\limits_{q^\prime=0}^{q-1}
[q-q^\prime]e^{2i(q-q^\prime)k_Fd} \\ 
\ \\
\times \eta(q-q^\prime)\int\limits_{0}^{\cal T} \frac{dt}{\cal T} \bigg\{\delta_{\alpha\beta}\sum\limits_{\gamma=1}^{N_r} {\cal S}^{(q,\tau)}_{\alpha\gamma}(t,\mu) {\cal S}^{(q^\prime,\tau)\star}_{\alpha\gamma}(t,\mu) \\
\ \\
\!\!\!\!- {\cal S}^{(q,\tau)}_{\alpha\beta}(t,\mu) {\cal S}^{(q^\prime,\tau)\star}_{\alpha\beta}(t,\mu)
- {\cal S}^{(q,\tau)}_{\beta\alpha}(t,\mu) {\cal S}^{(q^\prime,\tau)\star}_{\beta\alpha}(t,\mu) \bigg\}\,.
\end{array}
\end{equation}
\end{subequations}
\ \\
The function $\eta(q-q')$ is defined in Eq.\,(\ref{Eqdc_4E}).

At high temperatures the thermal noise of a driven system depends on the matrix $\hat g$ in the same manner as the thermal noise of a stationary scatterer depends on the conductance matrix. \cite{BB00}
Therefore, we can treat $\hat g$ as an effective (high-temperature) conductance matrix of a driven scatterer.

At adiabatic driving, ${\cal T}\gg\tau_0$,  the matrix $\hat g^{(ad)}(t)\equiv \lim\limits_{\Omega\to 0}\hat g(t)$ is independent of $\Omega$. 
It can be expressed in terms of reflection/transmission coefficients for point-like scatterers. 
If $\hat L$ and $\hat R$ are symmetric in lead indices then we get,

\begin{equation}
\label{Eqzfn_5}
\frac{1}{g_{\alpha\ne\beta}^{(ad)}(t)} = 
\frac{1}{T_L(t)} + \frac{1}{T_R(t)} -1 \,.
\end{equation}
\ \\
Here 
$T_{L} = |L_{12}|^2$,
$T_R = |R_{12}|^2$, and 
$g_{\alpha\alpha}^{(ad)}(t) = 1 - g_{\alpha\ne\beta}^{(ad)}(t)$.
Note that the effective transmission coefficient $g_{12}^{(ad)}(t)$ differs from the probability for incoherent (sequential) tunneling through two barriers, $g^{(seq)}_{12}(t) = \left[T_L^{-1}(t) + T_R^{-1}(t) \right]^{-1}$. 
The latter is attributed to structures with inelastic relaxation processes inside the well \cite{Buttiker86} whereas Eq. (\ref{Eqzfn_5}) 
can be obtained as a consequence of incoherent but energy conserving scattering \cite{deJong96} in the well.
Note also, that in Ref.\,\onlinecite{MB01} it was shown that in the adiabatic regime the dc current pumped by the double barrier structure with strong inelastic scattering is determined by the sequential conductance $g^{(seq)}_{12}$. 
In contrast the high temperature pumped current $I^{(0)}$, Eq.\,(\ref{Eqdc_4B}), at $\Omega\to 0$ is determined by the quasi-elastic adiabatic conductance $g^{(ad)}_{12}$.

\begin{figure}[t]
  \vspace{0mm}
  \centerline{
   \epsfxsize 8cm
    \epsffile{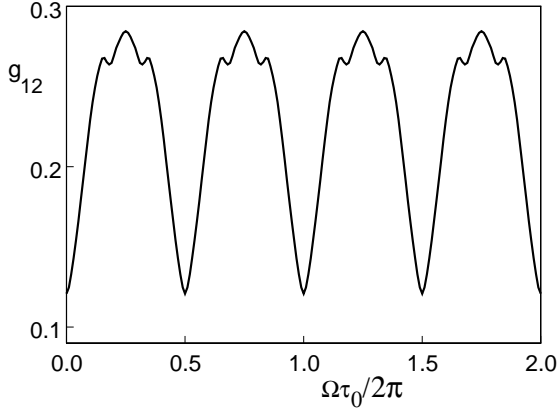}
             }
  \vspace{0mm}
  \nopagebreak
  \caption{The effective transmission probability $g_{12}$ as a function of $\Omega$. The parameters are the same as in Fig.\,\ref{fig2}.}
\label{fig3}
\end{figure}

At high driving frequencies $\hat g$ becomes dependent on $\Omega$. 
In Fig.\,\ref{fig3} we give the transmission probability $g_{12}$ as a function of the non-adiabaticity parameter $\Omega\tau_0$.

\subsection{Shot noise}

Similar to the thermal noise, the shot noise has a part which vanishes at $T\gg T^\star$.
To show this we use Eq.\,(\ref{Eqdb_13A}) and represent the two-particle scattering matrix $\hat{\it \Sigma}_{in}$, Eq.\,(\ref{Eqzfn_3}), as follows:

\begin{equation}
\label{Eqzfn_6}
\begin{array}{l}
\!{\it \Sigma}_{in,\alpha\beta}(t_1,t_2;E) = 
\sum\limits_{\gamma}
\sum\limits_{q=-\infty}^{\infty}
e^{2iQ_{\alpha\beta;\gamma}^{(q)}kd}
{\it \Sigma}_{\alpha\beta;\gamma}^{(q)}(t_1,t_2;E) , \\
\ \\
Q_{\alpha\beta;\gamma}^{(q)} = q+\frac{\delta_{\beta\gamma}-\delta_{\alpha\gamma}}{2}, \\
\ \\
{\it \Sigma}_{\alpha\beta;\gamma}^{(q\ge 0)}(t_1,t_2;E) = 
\sum\limits_{r=0}^{\infty}
{\cal S}_{\alpha\gamma}^{(r + q,\tau)}(t_1,E)
{\cal S}_{\beta\gamma}^{(r,\tau)\star}(t_2,E)\,, \\
\ \\
{\it \Sigma}_{\alpha\beta;\gamma}^{(q< 0)}(t_1,t_2;E) =\sum\limits_{r=0}^{\infty}
{\cal S}_{\alpha\gamma}^{(r,\tau)}(t_1,E)
{\cal S}_{\beta\gamma}^{(r - q,\tau)\star}(t_2,E)\,, \\
\end{array}
\end{equation}

Substituting this equation into Eq.\,(\ref{Eqzfn_2B}) and integrating over energy we obtain the following:

\begin{equation}
\label{Eqzfn_7}
\begin{array}{l}
\!{\bf P}_{\alpha\beta}^{(sh)} = \frac{2e^2}{h}
\sum\limits_{q_1,q_2=-\infty}^{\infty}
\sum\limits_{\gamma,\delta}
e^{2ik_Fd\Theta_{\gamma\delta}^{(-)}[q_1,q_2]}
{\cal P}_{\gamma\delta}^{(sh)}(q_1,q_2)\,, \\
\ \\
{\cal P}_{\gamma\delta}^{(sh)} = 
\int\limits\!\!\! \int\limits_{0}^{\cal T} \frac{dt_1d t_2}{2{\cal T}^2} 
\sum\limits_{m=-\infty}^{\infty} e^{im\left(t_2 - t_1 + \Omega\tau_0 \Theta_{\gamma\delta}^{(+)}[q_1,q_2]
\right)}
 \\
\ \\
\hspace{10mm}\times
{\cal F}_{\gamma\delta}^{(q_1q_2)}(m)
{\it \Sigma}_{\alpha\beta;\gamma}^{(q_1)\star}(t_1,t_2;\mu)
{\it \Sigma}_{\alpha\beta;\delta}^{(q_2)}(t_1,t_2;\mu)\,, 
\end{array}
\end{equation}
\ \\
where 
$$
\begin{array}{l}
\Theta_{\gamma\delta}^{(\pm)}[q_1,q_2] = Q_{\alpha\beta;\delta}^{(q_2)} \pm Q_{\alpha\beta;\gamma}^{(q_1)}\,, \\
\ \\
{\cal F}_{\gamma\delta}^{(q_1q_2)} = \eta\left(\Theta_{\gamma\delta}^{(-)}[q_1,q_2]\right)
\bigg\{ - 2k_BT\cos\left(m\Omega\tau_0\Theta_{\gamma\delta}^{(-)}[q_1,q_2]\right) \\
\ \\
\hspace{15mm}+\hbar\coth\left(m\frac{\hbar\Omega}{2k_BT}\right)\frac{\sin\left(m\Omega\tau_0\Theta_{\gamma\delta}^{(-)}[q_1,q_2]\right)}{\tau_0\Theta_{\gamma\delta}^{(-)}[q_1,q_2]}
\bigg\}\,.
\end{array}
$$

At high temperatures, $T\gg T^\star\sim\tau_0^{-1}$, only terms with $Q_{\alpha\beta;\delta}^{(q_2)} = Q_{\alpha\beta;\gamma}^{(q_1)}$ contribute to the shot noise. 
In this limit, for instance, the auto-correlations are:

\begin{equation}
\label{Eqzfn_8}
\begin{array}{c}
{\bf P}_{\alpha\alpha}^{(sh,0)} = \frac{2e^2}{h}
\int\limits\!\!\! \int\limits_{0}^{\cal T} \frac{dt_1 dt_2}{{\cal T}^2} 
\sum\limits_{q=0}^{\infty}
F_T(t_2 - t_1 + 2q\Omega\tau_0) \\
\times
(2-\delta_{q0})\left|\sum\limits_{\gamma}
{\it{\it \Sigma}}_{\alpha\alpha;\gamma}^{(q)}(t_1,t_2;\mu)\right|^2\,, \\
\ \\
\!\!F_T(t) = 2k_BT\sum\limits_{m=1}^{\infty}\cos(mt) 
\left\{\frac{m\hbar\Omega}{2k_BT}\coth\left(\frac{m\hbar\Omega}{2k_BT}\right) - 1\right\}\,.
\end{array}
\end{equation}

From Eq.\,(\ref{Eqzfn_8}) it follows that for $\hbar\Omega\ll k_BT$ the shot noise is inversely proportional to the temperature: ${\bf P}^{(sh,0)}\sim (k_BT)^{-1}$. 
The dependence on frequency $\Omega$ is more subtle. 
For adiabatic driving $\Omega\ll\tau_0^{-1}$ only $F_T$ depends on the driving frequency, $F_T\sim \Omega^2$. 
Therefore, in this case we get ${\bf P}^{(sh,0)}\sim \Omega^2/(k_BT)$ in accordance with Refs.~\onlinecite{AEGS01,MBnoise04}.
In contrast, for non-adiabatic driving, $\tau_0^{-1}\ll\Omega\ll k_B T/\hbar$, the matrix $\hat S_{in}$ and hence $\hat {\it \Sigma}$ oscillates in $\Omega$. 
As a consequence the shot noise oscillates in $\Omega$ with increasing ($\sim \Omega^2$) amplitude. 
Since $\hbar\Omega\ll k_BT$ the thermal noise dominates over the shot noise in both the adiabatic and the non-adiabatic regimes.

The situation is different in the strongly non-adiabatic regime, $\tau_0^{-1}\ll k_BT\ll\hbar\Omega$.
In this case $F_T\sim\Omega$.
Therefore, the shot noise is temperature independent and it oscillates in frequency with amplitude growing linearly in $\Omega$. 
In this regime the shot noise exceeds the thermal noise.

\section{The effect of an external bias}
\label{bias}

In this section we extend the formalism presented above to the case when there is in addition an external bias. 
Several factors make such a consideration important. \cite{MBac04}
First, usually the mesoscopic structure is a part of some electrical circuit with non-zero impedance used to measure a current. 
In the presence of a pump current the voltage arising at the external impedance acts back on the sample.
Second, the gates with oscillating voltages used to drive the sample can produce oscillating voltages at all the terminals. \cite{DMH03}
The rectification of ac currents due to external ac voltages by the oscillating scatterer \cite{Brouwer01,PB01,DMH03,MP07} can lead to a dc current in addition to a pumped current.

\subsection{The effect of a dc bias}

To be definite we assume that the dc voltage $eV$ is applied to the right reservoir:

\begin{equation}
\label{Eqb_1}
\mu_1 = \mu\,, \quad \mu_2 = \mu + eV\,, \quad eV \ll \mu\,.
\end{equation}

We calculate the dc current, $I_{1}$,  via Eq.\,(\ref{Eqdc_1}).
In the high-temperature limit, $T\gg T^{\star}$ we find: $I_{1} = I_{1}^{(0)} + I_{1}^{(dc)}$.
Here $I_{1}^{(0)}$ is the pumped current, Eq.\,(\ref{Eqdc_4A}), and $I_{1}^{(dc)}$ is a current due to a dc voltage:

\begin{equation}
\label{Eqb_2}
I_{1}^{(dc)} = VG_0 g_{12}\,, \quad\quad T\gg T^{\star}\,,
\end{equation}
\ \\
where $G_0 = e^2/h$ is the conductance quantum (per single spin channel), 
the effective conductance $g_{12}$ is defined in Eq.\,(\ref{Eqzfn_4A}).
Note that the pumped current and the dc voltage-driven current are conserved separately, $\sum_\alpha I_{\alpha}^{(0)} = 0$ and $\sum_\alpha I_{\alpha}^{(dc)} = 0$.

In the two terminal case of interest here, the last equation leads to $g_{12} = g_{21}$.
In addition using Eqs.\,(\ref{Eqzfn_4A}) and (\ref{m8_3_A}) 
we find that the sum over all the incoming leads equals to unity, $\sum_\beta g_{\alpha\beta}(t) = 1$.
This condition allows us to interpret $g_{\alpha\beta}(t)$ as equal-arrival-time  transmission (for $\alpha\ne\beta$) or reflection (for $\alpha=\beta$) probabilities for the particles with Fermi energy. 
In other words, the quantity $g_{\alpha\beta}(t)$ in Eq.\,(\ref{Eqzfn_4A}), is a probability for an electron entering the system through the lead $\beta$ to leave the system at a given time moment $t$ through the lead $\alpha$.
This probability is a sum of probabilities, $\left|{\cal S}_{\alpha\beta}^{(q,\tau_{0})}(t,\mu) \right|^2$, corresponding to electrons entering the system at time moments $t - \tau_{0}(2q + 1 - \delta_{\alpha\beta})$.
We emphasize that this interpretation applies at high temperatures, $T\gg T^{\star}$. 
At lower temperatures only the relation Eq.\,(\ref{Eqdb_15}), averaged over a pumping cycle, holds.

In the non-adiabatic, high-temperature regime, $\Omega\tau_0\gg 1$ and $T\gg T^{\star}$, both the pumped current $I^{(0)}$ and the dc voltage-driven current $I^{(dc)}$ oscillate at the driving frequency (see Fig.\,\ref{fig2} and Fig.\,\ref{fig3}) with the same period.
The important difference is that the amplitude of oscillations of  $I^{(0)}$ grows linearly with $\Omega$.
The pumped current can be expressed in terms of the average charge transferred per cycle $Q(\Omega)$ multiplied by the frequency, $I^{(0)} = Q \Omega/(2\pi)$. The charge $Q$ pumped per cycle as a function of the pump frequency $\Omega$ is given in Fig.\,\ref{fig2}.
The amplitude of oscillations of $I^{(dc)}$ is linear in the  dc voltage $V$.
Therefore, varying both $\Omega$ and $V$ one can, in principle, distinguish between the pumped current and the external voltage-driven current.

We reemphasize that the quantities $g_{\alpha\beta}$ define both the thermal noise and the dc conductance of a driven scatterer. 
In contrast the rectification of currents due to external ac voltages is defined by different quantities.

\subsection{The effect of an ac bias}

Let us suppose that ac voltages are applied to the contacts of a dynamic scatterer ($\alpha = 1, 2, \cdots, N_r$): 

\begin{equation}
\label{Eqb_3}
V_{\alpha}(t) = V_{0,\alpha} + V_{\alpha}\cos(\Omega t + \varphi_{V_{\alpha}}), 
~~eV_{0, \alpha},eV_{\alpha} \ll \mu_{\alpha}\, . 
\end{equation}
\ \\
Then the dc current flowing through lead $\alpha$ is \cite{MBac04}

\begin{equation}
\label{Eqb_4}
\begin{array}{c}
\tilde I_{\alpha} = \frac{e}{h} \int\limits_{0}^{\infty} dE  \Bigg\{ - f_{\alpha}(E) + 
\sum\limits_{\beta}  
\sum\limits_{n=-\infty}^{\infty}  
f_{\beta}(E - n\hbar\Omega)  \\
\ \\
\!\!\!\!\times
\sum\limits_{m,q=-\infty}^{\infty}  
S^{*}_{F,\alpha\beta}(E,E_q) S_{F,\alpha\beta}(E,E_m)  
\Upsilon_{\beta,n+q}^{\star}\Upsilon_{\beta,n+m}
\Bigg\}\,. 
\end{array}
\end{equation}
\ \\
Here $\Upsilon_{\beta,n}$ is the Fourier coefficient for the function $\Upsilon_{\beta}(t)$. The latter is given by Eq.\,(\ref{Eqdb_4}) with $eU$ and $\varphi_U$ being replaced by $eV_{\beta}$ and $\varphi_{V_{\beta}}$, respectively.
We use the symbol tilde to indicate that the dc current $\tilde I_{\alpha}$ is calculated in the presence of external ac voltages.
The Fermi functions $f_{\alpha(\beta)}$ entering Eq.\,(\ref{Eqb_4}) depend on the corresponding chemical potentials $\mu_{\alpha(\beta)} = \mu + eV_{0,\alpha(\beta)}$.

To simplify the equation given above, we introduce the matrix $\hat S_{out}(E,t)$ whose Fourier coefficients define the elements of the Floquet scattering matrix in the following way: \cite{MBac04}

\begin{equation}
\label{Eqb_5}
\hat S_{out,n}(E) = \hat S_{F}(E,E_{-n})\,.
\end{equation}
\ \\
The lower index "out" indicates that the matrix introduced in this way depends on an outgoing energy entering the corresponding element of the Floquet scattering matrix. In contrast the matrix $\hat S_{in}$ which we introduced earlier is a function of an incoming energy.

Substituting Eq.\,(\ref{Eqb_5}) into Eq.\,(\ref{Eq_new2B}) we derive the following constraint to the matrix $\hat S_{out}(E,t)$ [compare with Eq.\,(\ref{Eqdb_15})]: 

\begin{equation}
\label{Eqb_6}
\int\limits_{0}^{\cal T} \frac{dt}{{\cal T}} 
\hat S_{out}(E,t) \hat S_{out}^{\dagger}(t,E) = \hat I\, .
\end{equation}
\ \\
To calculate this scattering matrix we find $\hat S_{F}(E_n,E)$ using Eqs.\,(\ref{Eq_new1}), (\ref{Eqdb_13A}) and (\ref{Eqdb_13}). 
Then with the help of Eqs.\,(\ref{Eqb_5}) after the inverse Fourier transformation we find:

\begin{equation}
\label{Eqb_7}
\begin{array}{c}
S_{out,\alpha\beta}(E,t)=\sum\limits_{q=0}^{\infty} e^{2iq_{\alpha\beta}kd}  \tilde{\cal S}_{\alpha\beta}^{(q,\tau)}(E,t)\,, \\
\ \\
\tilde{\cal S}_{\alpha\beta}^{(q,\tau)}(E,t) = {\cal S}_{\alpha\beta}^{(q,\tau)}(t+2q_{\alpha\beta}\tau,E)\, .
\end{array}
\end{equation}
\ \\
The amplitudes ${\cal S}_{\alpha\beta}^{(q,\tau)}(t,E)$ are defined in Eq.\,(\ref{Eqdb_13}).
We interpret $S_{out,\alpha\beta}(E,t)$ as an amplitude corresponding to scattering of an electron which enters the system through the lead $\beta$ at time moment $t$  and leaves the system through the lead $\alpha$ with energy $E$. 
Since the energy of the outgoing state is fixed, the time when the particle leaves the system  is not fixed. 
Correspondingly, since the time moment when the particle enters the system is fixed, the incoming energy is not fixed.

Now we proceed with the calculation of the current.
After simple transformations we rewrite Eq.\,(\ref{Eqb_4}) in the following way:

\begin{equation}
\label{Eqb_8}
\tilde I_{\alpha} = I_{\alpha} + I_{\alpha}^{(ex)}\,,
\end{equation}
\ \\
where the current $I_{\alpha}$ is the pumped current, Eq.\,(\ref{Eqdc_1}), in the absence of the oscillating voltages at contacts. 
The excess current $I_{\alpha}^{(ex)}$ due to ac voltages is

\begin{equation}
\label{Eqb_9}
\begin{array}{c}
I_{\alpha}^{(ex)} = \frac{e}{h}\int dE 
\sum\limits_{\beta}
\sum\limits_{n=-\infty}^{\infty}
\big[f_{\beta}(E_{-n}) - f_{\alpha}(E)\big] \\
\ \\
\times
\left\{
\left| \left(S_{out,\alpha\beta}\Upsilon_{\beta} \right)_{n} \right|^2 - 
\left|S_{out,\alpha\beta,n}\right|^2  
\right\}\,.
\end{array}
\end{equation}
\ \\
The effect of oscillating potentials at external reservoirs is formally only a phase shift [see Eq.\,(\ref{Eqdb_4})] of the corresponding elements of the scattering matrix $\hat S_{out}$.
However due to the directional asymmetry of time-dependent scattering, $S_{out,\alpha\beta}\ne S_{out,\beta\alpha}$, the current $I_{\alpha}^{(ex)}$ can be non-zero even if the voltages at all the contacts are the same, $V_{\alpha}(t) = V(t)$.
Therefore, in accordance with the conclusion reached in Ref.\,\onlinecite{MBac04} the excess current contains two generic contributions, $I_{\alpha}^{(ex)}=I_{\alpha}^{(rect)}+I_{\alpha}^{(int)}$.
The first one, $I_{\alpha}^{(rect)}$, results from rectification of ac currents due to external ac voltages by the oscillating scatterer.
The second one, $I_{\alpha}^{(int)}$, is due to interference between internal and external ac currents. 
On the other hand one can think of the latter contribution as due to external voltages which act as additional pump parameters. 
Note that the interference contribution was also addressed in Ref.\,\onlinecite{BB08}.

In the high-temperature limit Eq.\,(\ref{Eqb_9}) can be simplified.
Using Eq.\,(\ref{Eqb_7}) and performing integration over energy in the same way as we did to get Eq.\,(\ref{Eqdc_4}), we obtain:

\begin{subequations}
\label{Eqb_10}
\begin{equation}
\label{Eqb_10A}
I_{\alpha}^{(ex)} = G_0\int\limits_{0}^{\cal T}\frac{dt}{\cal T}
\sum\limits_{\beta}V_{\beta}(t)\tilde g_{\alpha\beta}(t) , \quad T\gg T^{\star}\,,
\end{equation}

\begin{equation}
\label{Eqb_10B}
\tilde g_{\alpha\beta}(t) = 
\sum\limits_{q=0}^{\infty}
\left|\tilde S_{\alpha\beta}^{(q,\tau_0)}(t,\mu) \right|^2\, .
\end{equation}
\end{subequations}

The pumped current $I_{\alpha}$, Eq.\,(\ref{Eqdc_1}), is conserved: $\sum_{\alpha} I_{\alpha} = 0$.
Therefore, the conservation of the whole current $\tilde{I}_{\alpha}$, Eq.\,(\ref{Eqb_8}), implies $\sum_\alpha I_{\alpha}^{(ex)} = 0$.
The latter in turn leads to the following constraint: $\sum_{\alpha} \tilde g_{\alpha\beta}(t) = const$. 
Quite similar to the case with $g_{\alpha\beta}$ we find that this constant is the unity.
Using this condition one can split $I_{\alpha}^{(ex)}$ into the rectification [i.e., vanishing at $V_1(t) = V_2(t)$] and interference parts as follows:

\begin{figure}[t]
  \vspace{0mm}
  \centerline{
   \epsfxsize 8cm
     \epsffile{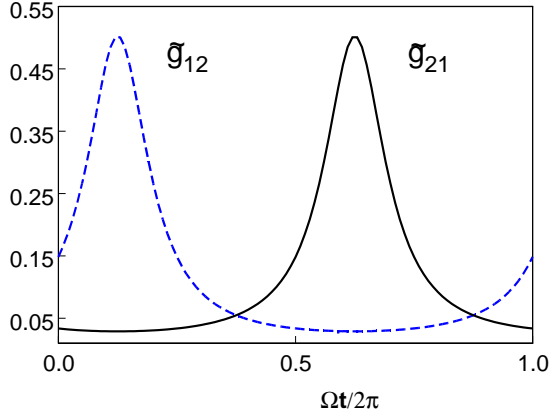}
            }
 \vspace{0mm}
 \nopagebreak
  \caption{(color online) The transmission probabilities $\tilde g_{21}(t)$ (black solid line) and $\tilde g_{12}(t)$ (blue dashed line) are given as a function of time for one pumping period. The non-adiabaticity parameter is $\Omega\tau_0=\pi$. Other parameters are the same as in Fig.\,\ref{fig2}. }
 \label{fig4}
\end{figure}

\begin{subequations}
\label{Eqb_10_1}
\begin{equation}
\label{Eqb_10_1A}
I_{1}^{(rect)} = G_0\int\limits_{0}^{\cal T}\frac{dt}{\cal T}
\big[V_{2}(t) - V_{1}(t)\big]\frac{\tilde g_{12}(t) + \tilde g_{21}(t)}{2}\, , 
\end{equation}

\begin{equation}
\label{Eqb_10_1B}
I_{1}^{(int)} = G_0\int\limits_{0}^{\cal T}\frac{dt}{\cal T}
\frac{V_{2}(t) + V_{1}(t)}{2}
\big[\tilde g_{12}(t) - \tilde g_{21}(t)\big]\,.
\end{equation}
\end{subequations}
\ \\
The interference contribution can be non-zero, $I_{1}^{(int)}\ne 0$, if  and only if the dynamical scatterer shows different time-dependent transmissions to the left and to the right, $\tilde g_{12}(t) \ne \tilde g_{21}(t)$, see Fig.\,\ref{fig4}. 
In addition, the time-dependent voltages should be present, $V_{\alpha}(t)\ne const$. 
The last statement follows from the fact that after averaging over time the matrix $\hat{\tilde g}$ becomes symmetric in lead indices.
Therefore, we conclude that the current $I_{1}^{(int)}$ is due to an interplay of the external induced dynamics and the internally induced dynamics.

\begin{figure}[t]
  \vspace{0mm}
  \centerline{
   \epsfxsize 8cm
   \epsffile{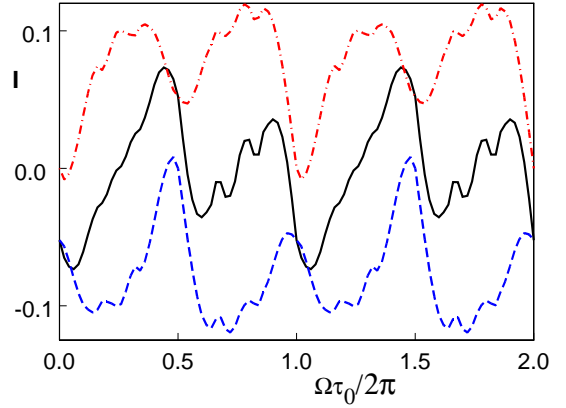}
            }
 \vspace{0mm}
 \nopagebreak
  \caption{(color online) The excess $I_{1}^{(ex)}$ (black solid line), rectified $I_{1}^{(rect)}$ (blue dashed line), and interference, $I_{1}^{(int)}$ (red dash-dotted line)  currents in units of $I_0 = VG_0$ as a function of $\Omega$. The external voltages are $V_2 = V\cos(\Omega t)$, $V_1 = 0$. Other parameters are the same as in Fig.\,\ref{fig2}. }
 \label{fig5}
\end{figure}

The excess current $I_{\alpha}^{(ex)}$, arising in the presence of external ac voltages, comprises two contributions: a pure rectification current, $I_{\alpha}^{(rect)}$, and an additional pumped current, $I_{\alpha}^{(int)}$. These two contributions differ from each other. In addition they differ from the current pumped by the unbiased dynamical scatterer. Formally, the last mentioned difference is due to the fact that $I_{\alpha}^{(ex)}$ is defined by the scattering matrix $\hat S_{out}$ while $I_{\alpha}$ is defined by the matrix $\hat S_{in}$. This results in a different frequency dependence,  see Fig.\,\ref{fig5} for $I_{\alpha}^{(ex)}$ and Fig.\,\ref{fig2} (black solid line) for the pumped current. 
Everything mentioned makes it difficult to extract a pure pumped current from an experimentally measured dc current.

If the voltages at contacts are time-independent, $V_{\beta}(t) = const$, equation (\ref{Eqb_10A}) transforms into Eq.\,(\ref{Eqb_2}) since the average over a pump period  for quantities $\tilde g_{\alpha\beta}$ and $g_{\alpha\beta}$ is the same. 
This follows directly from Eq.\,(\ref{Eqb_7}) (the second line).

Note by analogy with $g_{\alpha\beta}(t)$ we interpret $\tilde g_{\alpha\beta}(t)$ as equal-departure-time  probabilities for transmission/reflection of electrons with Fermi energy:
This quantity is the probability for an electron entering the system at a given [defined by the potential $V_{\beta}(t)$ in Eq.\,(\ref{Eqb_10A})] time moment $t$ through the lead $\beta$ to leave the system through the lead $\alpha$. 
Accordingly to Eq.\,(\ref{Eqb_10B}) the events with electrons leaving the system at different time moments do contribute to $\tilde g_{\alpha\beta}(t)$.
We note the equation (\ref{Eqb_10}) was obtained for high temperatures.
For lower temperature we have no instant-time quantity like $\tilde g_{\alpha\beta}(t)$ with a transparent physical meaning. 
The matrix $\hat S_{out}$ entering Eq.\,(\ref{Eqb_9}) obeys a constraint that is non-local in time, Eq.\,(\ref{Eqb_6}).

\section{Discussion and conclusion}
\label{concl}

We have analyzed current and noise generated by a dynamical double-barrier structure with and without an external bias at arbitrary temperature and at arbitrary frequency and amplitude of a periodic harmonic drive. 

To perform such an analysis we have developed a method to calculate analytically the scattering matrices $\hat S_{in}(t,E)$ and $\hat S_{out}(E,t)$ whose Fourier coefficients define the elements of the Floquet scattering matrix, Eqs.\,(\ref{Eq_new1}) and (\ref{Eqb_5}).
For the elements of these matrices which we found for the double-barrier structure 
there exists a simple interpretation. 
This interpretation allows to find the corresponding matrices for more involved systems.  
For instance, the matrix element $S_{in,\alpha\beta}(t,E) =  \sum_q e^{2iq_{\alpha\beta}kd}  {\cal S}_{\alpha\beta}^{(q,\tau)}(t,E)$, Eq.\,(\ref{Eqdb_13A}), can be calculated as a sum over all the possible paths which an electron can follow to propagating through the system. 
Here the partial amplitude $e^{2iq_{\alpha\beta}kd}  {\cal S}_{\alpha\beta}^{(q,\tau)}(t,E)$ corresponds to a particle entering the system with energy $E$ through the lead $\beta$ and leaving the system through the lead $\alpha$ at a time moment $t$. 
The summation index $q$ corresponds to the path ${\cal L}_{\alpha\beta}^{(q)}$ having definite length (in our case for $q>0$ this length is a product of the number $2q-\delta_{\alpha\beta}$ of reflections inside the system and the distance $d$ between the two barriers). 
Each partial amplitude is a product of amplitudes corresponding to an instantaneous interaction with time-dependent point-like barriers (in our case these amplitudes are the elements of matrices $\hat L$ and $\hat R$) and a number of amplitudes $e^{i[kd- e\hbar^{-1}\int^{t_j}_{t_j-\tau} dt^\prime U(t^{\prime})]}$ corresponding to free propagation (having a duration of $\tau=d/v$) between the barriers. 
The time moments, at which the corresponding amplitudes are calculated, are counted backwards along the  path in a descending order starting from the time moment $t$ when the particle leaves the system. 

We found an analogous interpretation for elements of the matrix $\hat S_{out}(E,t)$. 
The in- and out- scattering matrices become equal (and equal to the frozen scattering matrix) in the limit of a slow drive, $\Omega\to 0$.
At any finite driving frequencies ($\Omega\ne 0$) they are different. Nevertheless there is a simple relation between these matrices. To find it we use the micro reversibility of the equations of motion.
Let the driving parameters depend on time in the following way: $p_i(t) \sim \cos(\Omega t + \varphi_i)$, see, Eq.\,(\ref{Eqdb_1}). 
Then the elements of the Floquet scattering matrix are subject to the following symmetry: \cite{MBmag05}

\begin{equation}
\label{Eqc_1}
S_{F,\alpha\beta}(E, E_{n}; H, \varphi) = S_{F,\beta\alpha}(E_n, E; -H, -\varphi)\,.
\end{equation}
\ \\
Here $\varphi$ is the set of all the $\varphi_{i}$'s. 
In addition we introduced a possibly present magnetic field $H$.
Substituting Eqs.\,(\ref{Eqdb_9}) and (\ref{Eqb_5}) into Eq.\,(\ref{Eqc_1}) and performing the inverse Fourier transformation we arrive at the desired relation:

\begin{equation}
\label{Eqc_2}
S_{in,\alpha\beta}(t,E; H, \varphi) = S_{out,\beta\alpha}(E,-t; -H, -\varphi)\,.
\end{equation}
\ \\
Note that each term of the decompositions, Eq.\,(\ref{Eqdb_13A}) and Eq.\,(\ref{Eqb_7}), satisfies Eq.\,(\ref{Eqc_2}) separately.

We showed that the scattering matrix $\hat S_{in}$ defines currents generated by the oscillating scatterer itself, while the dual scattering matrix $\hat S_{out}$ defines currents flowing through the scatterer under an external bias.
Here we calculated the matrix $\hat S_{in}$ for arbitrary frequency and consequently we can calculate the current in the adiabatic as well as in the non-adiabatic regime.

The adiabatic regime can be defined \cite{MBstrong02} by considering the energy dependence of the stationary scattering matrix . Since both transmission and reflection times are of the 
order of the dwell time, the latter is (at least in the absence of interactions) the relevant time scale.
The results obtained here allows us to conclude that the dwell time is appropriate to define the adiabatic regime for zero-temperature pumping. In contrast, 
at high temperatures another time, the propagation time $\tau_0=d/v$ (necessary for an electron to travel between the two barriers separated by the distance $d$), defines a relevant time scale. 

Let us consider the scattering processes which are relevant for pumping.
Generally speaking the interplay between the quantized energy exchange and the interference of photon-assisted scattering amplitudes is at the origin of the pump effect. \cite{BM05}
As we showed, there are two types of interference processes. 
The first one takes place within the same spatial path (intra-channel interference).
A second process includes the interference between different spatial paths (inter-channel interference).
The former contributes to the generated current at low as well as at high temperatures, 
while the latter contributes only to the low temperature pump effect.
The dwell time (defined in the stationary case) depends essentially on interference processes of the second type (there it nothing to interfere within the same spatial path in the stationary case).
Therefore, we expect that the dwell time is relevant only to low temperature pumping. In contrast 
for high temperatures we expect that a time of the order of the propagation time $\tau_0$ defines the lower bound for the period of a drive separating adiabatic and non-adiabatic dynamical regimes.

Our example suggest that distinction of inter- and intra-channel interference processes is also important for the magnetic field symmetry of the current pumped by a two-terminal scatterer.
The intra-channel interference processes are not affected by the magnetic field. 
However the inter-channel processes are sensitive to the magnetic field.
Therefore, we can conjecture that with increasing frequency and/or temperature (when only intra-channel interference processes matter) the pumped current has to become symmetric in magnetic field like the stationary (two-terminal) conductance. 
This is in striking contrast with the low temperature adiabatic pump effect which has no definite symmetry under the magnetic field reversal, see, e.g., Refs.~\onlinecite{Brouwer98,ZSA99}. 
Our conjecture agrees with experimental results presented in Refs.~\onlinecite{DMH03,VDM05} for large pumping frequencies and powers.

The method presented in the present paper is also useful to calculate the time-dependent current generated by the dynamical scatterer. 
Actually it was used in Ref.\,\onlinecite{MSB08} to describe the nonlinear response of a coherent capacitor to a large amplitude drive.  
In a recent experiment \cite{Feve07} it was demonstrated that such a capacitor can serve as a very fast single electron source.

In conclusion, we developed a method to calculate analytically the Floquet scattering matrix of a dynamical mesoscopic system consisting of a set of point-like scatterers and uniform potentials in between them. 
Our approach allows to go beyond the adiabatic approximation and to analyze the properties of the generated current and its noise at arbitrary frequency and amplitude of the drive.

\begin{acknowledgments}
This work was supported by the Swiss National Science Foundation. 
\end{acknowledgments}

\appendix

\section{The Floquet scattering matrix}
\label{appendix0}

Let us consider a one-dimensional structure consisting of two oscillating $\delta$-function barriers and an oscillating spatially uniform potential in between the barriers. 
All potentials oscillate with the same frequency. 
The single-particle scattering at a periodically driven system is described via the Floquet scattering matrix $\hat S_{F}$. \cite{MBstrong02, MBmag05}
To find its elements one needs to solve the Schr\"{o}dinger equation for an electron wave function $\Psi_{E}(t,x)$ with, in our case, the following time-dependent potential:

\begin{equation}
\label{Eqdb_1}
\begin{array}{l}
V(t,x) = V_L(t)\delta(x) + V_R(t)\delta(d-x) \\
\ \\
\hspace{3cm} + eU(t)\theta(x)\theta(d-x)\,, \\
\ \\
V_{L/R}(t) = V_{0L/R} + 2V_{1L/R}\cos(\Omega t+ \varphi_{L/R})\,, \\
\ \\
U(t) = U\cos(\Omega t+ \varphi_U)\,,
\end{array}
\end{equation}
\ \\
where $\theta(x)$ is the Heaviside step function.

To calculate $S_{F,11}$ and $S_{F,21}$ we consider a plain wave of a unit amplitude with energy $E$ incident from the left. 
Then the wave function $\Psi_{E}(t,x)$ outside the scatterer can be written as follows:

\begin{equation}
\label{Eqdb_2}
\begin{array}{cc}
\Psi_{E}(t,x) = e^{-i\frac{E}{\hbar}t}\sum\limits_{n=-\infty}^{\infty}
e^{-in\Omega t} \psi_{n}(x)\,, \\
\ \\
\!\!\!\!\!\!\psi_{n}(x) = \left\{
\!\begin{array}{cc}
\delta_{n0}e^{ikx} + \sqrt{\frac{k}{k_n}}S_{F,11}(E_n, E)e^{-ik_n x}\,, \!\!& x<0\,, \\
\ \\
\sqrt{\frac{k}{k_n}}S_{F,21}(E_n, E)e^{ik_n (x-d)}\,, \!\!& x>d\,.
\end{array}
\right.
\end{array}
\end{equation}
\ \\
Here 
$E_n = E + n\hbar\Omega$, 
$k_n = \sqrt{2m_eE_n}/\hbar$, 
with $m_e$ being the electron mass.

Following Ref.\,\onlinecite{Wagner94} we represent the functions $\psi_{n}(x)$ inside the scatterer in the form:

\begin{equation}
\label{Eqdb_3}
\psi_n(x) = \sum\limits_{l=-\infty}^{\infty} \Upsilon_{n-l}
\left(a_l e^{ik_l x} + b_l e^{-ik_l x} \right)\,,~~  0 < x < d\,, 
\end{equation}
\ \\
where the coefficients $a_l$, $b_l$ are independent of x and t.
Here $\Upsilon_{n}$ is the Fourier coefficient for a function $\Upsilon(t)$ describing the effect of an oscillating uniform potential $U(t)$,

\begin{equation} 
\label{Eqdb_4} 
\begin{array}{c}
\Upsilon_{n} = \int\limits_{0}^{\cal T} \frac{dt}{{\cal T}}
e^{in\Omega t}  \Upsilon(t)\,, \\
\ \\ 
\Upsilon(t) = \exp\left[- i \frac{eU}{\hbar\Omega}\sin(\Omega t + \varphi_U) \right]\,,
\end{array}
\end{equation} 
\ \\
where ${\cal T}=2\pi/\Omega$ is the period of the drive.

To match the wave function coefficients at different spatial regions we use the boundary conditions formulated in terms of the scattering matrices $\hat L_{F}$ and $\hat R_{F}$ for point-like scatterers located at $x=0$ and $x=d$, respectively.
These conditions are:

\begin{equation}
\label{Eqdb_5}
\begin{array}{l}
\sqrt{\frac{k}{k_n}}S_{F,11}(E_n, E) = 
\sqrt{\frac{k}{k_n}}L_{F,11}(E_n, E)  \\
\hspace{10mm}
+ \sum\limits_{m=-\infty}^{\infty}\sqrt{\frac{k_m}{k_n}}L_{F,12}(E_n, E_m)
\sum\limits_{l=-\infty}^{\infty}\Upsilon_{m-l} b_l\,,  \\
\ \\
\sum\limits_{p=-\infty}^{\infty}\Upsilon_{n-p} a_p = 
\sqrt{\frac{k}{k_n}}L_{F,21}(E_n, E) \\
\hspace{10mm}
+ \sum\limits_{m=-\infty}^{\infty}\sqrt{\frac{k_m}{k_n}}L_{F,22}(E_n, E_m)
\sum\limits_{l=-\infty}^{\infty}\Upsilon_{m-l} b_l\,,  \\
\ \\
\sum\limits_{p=-\infty}^{\infty}\Upsilon_{n-p} b_pe^{-ik_pd} = \\
\hspace{5mm}
\sum\limits_{m=-\infty}^{\infty}\sqrt{\frac{k_m}{k_n}}R_{F,11}(E_n, E_m)
\sum\limits_{l=-\infty}^{\infty}\Upsilon_{m-l} a_le^{ik_ld}\,,  \\
\ \\
\sqrt{\frac{k}{k_n}}S_{F,21}(E_n, E) = \\
\hspace{5mm}
\sum\limits_{m=-\infty}^{\infty}\sqrt{\frac{k_m}{k_n}}R_{F,21}(E_n, E_m)
\sum\limits_{l=-\infty}^{\infty}\Upsilon_{m-l} a_le^{ik_ld}\,.
\end{array}
\end{equation}

We suppose that the frequency $\Omega$ of the drive is small compared with the relevant electron energy $E$,

\begin{equation}
\label{Eqdb_6}
\hbar\Omega\ll E\,.
\end{equation}
\ \\
We solve Eq.\,(\ref{Eqdb_5}) to zeroth order in $\hbar\Omega/E$.
Within this accuracy we can simplify Eq.\,(\ref{Eqdb_5}).
First, the Floquet scattering matrix $\hat X_{F}(E_n, E)$ for a point-like scatterer (i.e., for a scatterer with a spatial extend much smaller than the de Broglie wave length for an electron with energy E) can be expressed in terms of the Fourier coefficients $\hat X_n(E)$ for a frozen scattering matrix $\hat X(t,E) = \hat X(t+{\cal T},E)$ (i.e., the stationary scattering matrix with a strength being dependent on time):
\cite{MBmag05}

\begin{equation}
\label{Eqdb_7}
\hat X_{F}(E_n, E_m) = \hat X_{n-m}(E)\,, \quad \hat X = \hat L\,, \hat R\,.
\end{equation}
\ \\
For a single $\delta$-function barrier $V(t,x) = V(t)\delta (x)$ the frozen scattering matrix is well known. Its elements are 
$X_{\alpha\beta}(t,E) = k/[k+iV(t)m_e/\hbar^2] - \delta_{\alpha\beta}$.

Second, we can put

\begin{equation}
\label{Eqdb_8}
\frac{k_m}{k_n} \approx 1\,, \quad e^{ik_ld} \approx e^{ikd}e^{il\Omega\tau}\,,
\end{equation}
\ \\
where $\tau = d/v$  is the time necessary for an electron with energy $E$ to propagate from one barrier to the other barrier.

Next we use that the quantities $a_l$'s and $b_l$'s are Fourier coefficients for some functions $a(t) = a(t+{\cal T})$ and $b(t) = b(t+{\cal T})$, periodic in time. 
In addition we introduce a matrix $\hat S_{in}(t,E)$ periodic in time whose Fourier coefficients $\hat S_{in,n}(E)$ define the Floquet scattering matrix $\hat S_{F}(E_n, E)$ for the whole structure as follows:

\begin{equation}
\label{Eqdb_9}
\hat S_{in,n}(E) = \hat S_{F}(E_n, E)\,.
\end{equation}
\ \\
The lower index "in" indicates that this matrix is a function of an incoming energy. 
With these definitions one can easily perform the inverse Fourier transformation of Eqs.\,(\ref{Eqdb_5}) and arrive at 

\begin{subequations}
\label{Eqdb_10}
\begin{equation}
\label{Eqdb_10A}
S_{in, 11}(t,E) = L_{11}(t,E) + L_{12}(t,E)\Upsilon(t)b(t)\,,
\end{equation}

\begin{equation}
\label{Eqdb_10B}
\Upsilon(t)a(t) = L_{21}(t,E) + L_{22}(t,E)\Upsilon(t)b(t)\,,
\end{equation}

\begin{equation}
\label{Eqdb_10C}
e^{-ikd}\Upsilon(t)b(t+\tau) = e^{ikd}R_{11}(t,E)\Upsilon(t)a(t-\tau)\,,
\end{equation}

\begin{equation}
\label{Eqdb_10D}
S_{in, 21}(t,E) =  e^{ikd}R_{21}(t,E)\Upsilon(t)a(t-\tau)\,,
\end{equation}
\end{subequations}

Substituting Eq.\,(\ref{Eqdb_10C}) into Eq.\,(\ref{Eqdb_10B}) we find the non-local in time equation for the function $a(t)$ (for brevity we suppress the argument $E$): 

\begin{equation}
\label{Eqdb_11}
a(t) = \Upsilon^{\star}(t)L_{21}(t) + e^{i2kd}L_{22}(t)R_{11}(t-\tau)a(t-2\tau)\,,
\end{equation}
\ \\
where the star denotes complex conjugation.
Since the coefficients entering the above equation with amplitudes smaller than unity, we can formally write down the solution in the form of an infinite series, 

\begin{equation}
\label{Eqdb_12}
\begin{array}{l}
a(t) = \sum\limits_{q=0}^{\infty} e^{i2qkd} \lambda^{(q)}(t)
\Upsilon^{\star}(t-2q\tau)L_{21}(t-2q\tau)\,, \\
\ \\
\lambda^{(q>0)}(t) = \prod\limits_{j=0}^{q-1} L_{22}(t_{2j})R_{11}(t_{2j}-\tau)\,.
\end{array}
\end{equation}
\ \\
Here $ t_{2j} = t - 2j\tau$. At $q=0$ we put $\lambda^{(0)}(t)=1$.

Substituting Eq.\,(\ref{Eqdb_12}) into Eqs.\,(\ref{Eqdb_10}) we obtain the matrix elements $S_{in, \alpha1}$, $\alpha=1,2$.
To find the matrix elements $S_{in, \alpha2}$ a problem with a plane wave of a unit amplitude coming from the right has to be considered. 
The solution to these two problems is written in Eq.\,(\ref{Eqdb_13A}) with 

\begin{subequations}
\label{Eqdb_13}
\begin{equation}
\label{Eqdb_13B}
{\cal S}^{(q,\tau)}_{\alpha\beta}(t,E) = e^{-i\Phi_{\alpha\beta}^{(q,\tau)}} {\sigma}^{(q,\tau)}_{\alpha\beta}(t,E)\,, 
\end{equation}
\ \\
where $\Phi_{\alpha\beta}^{(q,\tau)}=e\hbar^{-1}\int^{t}_{t - \tau 2q_{\alpha\beta}} dt^\prime U(t^{\prime})$  and

\begin{equation}
\label{Eqdb_13C}
\begin{array}{l}
{\sigma}_{11}^{(0,\tau)}(t) = L_{11}(t)\, , \\
\ \\
\!\!\!\!{\sigma}_{11}^{(q>0,\tau)}\!(t)\! =\! L_{12}(t)R_{11}(t-\tau)
L_{21}(t_{2q}) \lambda^{(q-1)}(t-2\tau) , 
\end{array}
\end{equation}

\begin{equation}
\label{Eqdb_13D}
{\sigma}_{21}^{(q,\tau)}(t) = R_{21}(t)
L_{21}(t_{2q+1}) \lambda^{(q)}(t-\tau)\, ,
\end{equation}

\begin{equation}
\label{Eqdb_13E}
{\sigma}_{12}^{(q,\tau)}(t) = L_{12}(t)
R_{12}(t_{2q+1}) \rho^{(q)}(t-\tau)\, ,
\end{equation}

\begin{equation}
\label{Eqdb_13F}
\begin{array}{l}
{\sigma}_{22}^{(0,\tau)}(t) = R_{22}(t)\, , \\
\ \\
\!\!\!\!{\sigma}_{22}^{(q>0,\tau)}(t)\! =\! R_{21}(t)L_{22}(t-\tau)
R_{12}(t_{2q}) \rho^{(q-1)}(t-2\tau) .
\end{array}
\end{equation}
\end{subequations}
\ \\
Here the function $\rho^{(q)}(t)$ is 

\begin{equation}
\label{Eqdb_14}
\begin{array}{l}
\rho^{(q>0)}(t) = \prod\limits_{j=0}^{q-1} R_{11}(t_{2j})L_{22}(t_{2j}-\tau)\,, \\
\ \\
\rho^{(0)} = 1\,.
\end{array}
\end{equation}

In the next section we consider in detail adiabatic (to first order in $\Omega$) asymptotics for the scattering matrix derived above.

\section{Adiabatic drive: $\Omega\tau\ll 1$}
\label{appendix}

In this Appendix we verify that for a slowly driven scatterer the matrix $\hat S_{in} = \hat S_{in}^{(1)} + {\cal O}(\Omega^2)$ can be represented through the frozen scattering matrix $\hat S$ as it was proposed in Refs.~\onlinecite{MBac04,MBmag05}:

\begin{equation}
\label{Eqad_1}
\hat S_{in}^{(1)}(t,E) = \hat S(t,E) + \frac{i\hbar}{2}
\frac{\partial^2\hat S}{\partial t\partial E} + \hbar\Omega\hat A(t,E)\,.  
\end{equation}
\ \\
We calculate the matrix $\hat A$ which is responsible for generating a directional asymmetry in scattering on an adiabatically driven scatterer. In turn, this asymmetry leads to an adiabatic quantum pump effect.
The matrix $\hat A$ satisfies \cite{MBac04}, 

\begin{subequations}
\label{Eqad_2} 
\begin{equation}
\label{Eqad_2A}
\hbar\Omega\left(\hat S^{\dagger}\hat A 
+ \hat A^{\dagger}\hat S\right)
= \frac{1}{2}{\cal P}\{\hat S^{\dagger};\hat S \}\,,
\end{equation}

\begin{equation}
\label{Eqad_2B}
{\cal P}\{\hat S^{\dagger};\hat S \} =
i\hbar \left( \frac{\partial \hat S^{\dagger}}{\partial t}
\frac{\partial \hat S}{\partial E} -
\frac{\partial \hat S^{\dagger}}{\partial E}
\frac{\partial \hat S}{\partial t}
\right)\,.
\end{equation}
\end{subequations}
\ \\
This equation results from the current conservation law up to linear in $\Omega$ terms. 

To find the matrix $\hat A$ we use the anzats Eq.\,(\ref{Eqad_1}) and consider the solution Eq.\,(\ref{Eqdb_13A}) in the limit of low driving frequency $\Omega\to 0$.
In this limit the delay time $\tau$ is small compared with a driving period ${\cal T} = 2\pi/\omega$. 
We use the ratio $\epsilon = \tau/{\cal T}$ as a small parameter.

\subsection{Zeroth order}

In zeroth order in $\epsilon$ the matrix $\hat S_{in}^{(0)}$ coincides with the frozen scattering matrix $\hat S$. 
The frozen scattering matrix is the stationary matrix with parameters being dependent on time.
To calculate the frozen scattering matrix one can use Eqs.\,(\ref{Eqdb_13A})  (\ref{Eqdb_13}) and neglect changes in all the quantities during the delay time $\tau$. 
Therefore, we have to put $t_{2q}\approx t$ in all the terms in Eqs.\,(\ref{Eqdb_13}) and use:

$$
\Phi_{\alpha\beta}^{(q,\tau)} \approx eU(t)\tau\hbar^{-1}(2q+1-\delta_{\alpha\beta}).
$$
\ \\
As a result we find $\hat S_{in}^{(0)}(t,E) = \hat S(t,E)$ with matrix elements:

\begin{equation}
\label{Eqad_3}
\begin{array}{l}
S_{\alpha\beta}(t,E) = \sum\limits_{q=0}^{\infty} S_{\alpha\beta}^{(q)}(t,E)\,, \\
\ \\
S_{\alpha\beta}^{(q)}(t,E) = e^{i(kd - eU(t)\tau/\hbar)(2q+1-\delta_{\alpha\beta})} \sigma_{\alpha\beta}^{(q,0)}(t,E)\, ,
\end{array}
\end{equation}
\ \\
where the matrix $\hat \sigma^{(q,\tau)}(t,E)$ is defined in Eq.\,(\ref{Eqdb_13}).

\subsection{First order}

To calculate the matrix $\hat S_{in}^{(1)}$ we expand Eq.\,(\ref{Eqdb_13}) in powers of $\tau$ and keep only the terms linear in $\tau$. 
Our aim is to calculate the irreducible part of $\hat S_{in}^{(1)}$, i.e. the matrix $\hat A$.
To this end we use Eq.\,(\ref{Eqad_1}) with the matrix $\hat S$ given by Eq.\,(\ref{Eqad_3}). 
We take into account that the frozen scattering matrix $\hat S$ depends on time $t$ through $U(t)$ and the matrices $\hat L$ and $\hat R$ of the point-like scatterers.
The energy dependence of this matrix is due to the phase factor $e^{2iqkd}$ only. 
We neglect the energy dependence of the matrices $\hat L$ and $\hat R$ and take them at $E=\mu$. We can do so since $\hat L$ and $\hat R$ change on the energy scale of the order of $\mu$ while in our problem the much smaller scales of order $\hbar\Omega$ and $k_BT$ are relevant, see, Eqs.\,(\ref{Eqdb_6}) and (\ref{Eqdc_2}).

After a straightforward calculation we find

\begin{subequations}
\label{Eqad_4}
\begin{equation}
\label{Eqad_4A}
\hbar\Omega A_{\alpha\beta}(t,E)=\sum\limits_{q=0}^{\infty} S_{\alpha\beta}^{(q)}(t,E) {\cal A}_{\alpha\beta}^{(q)}(t,\mu)\, ,
\end{equation}
\ \\
where
\begin{equation}
\label{Eqad_4B}
{\cal A}_{11}^{(q)} = \tau_{0}q\frac{\partial}{\partial t}\ln\left(\frac{L_{12}}{L_{21}} \right)\,,
\end{equation}

\begin{equation}
\label{Eqad_4C}
{\cal A}_{21}^{(q)} = -\frac{\tau_{0}(2q+1)}{2}\frac{\partial}{\partial t}\ln\left(\frac{L_{21}}{R_{21}} \right) - \frac{\tau_{0}q}{2}\frac{\partial}{\partial t}\ln\left(\frac{R_{11}}{L_{22}} \right)\,,
\end{equation}

\begin{equation}
\label{Eqad_4D}
{\cal A}_{12}^{(q)} =- \frac{\tau_{0}(2q+1)}{2}\frac{\partial}{\partial t}\ln\left(\frac{R_{12}}{L_{12}} \right) - \frac{\tau_{0}q}{2}\frac{\partial}{\partial t}\ln\left(\frac{L_{22}}{R_{11}} \right)\,,
\end{equation}

\begin{equation}
\label{Eqad_4E}
{\cal A}_{22}^{(q)} = \tau_{0}q\frac{\partial}{\partial t}\ln\left(\frac{R_{21}}{R_{12}} \right)\,.
\end{equation}
\end{subequations}
\ \\
At $U(t)=0$, the above expressions coincide with those obtained in Ref.\,\onlinecite{MBmag05}.

Equation (\ref{Eqad_4}) shows that the matrix $\hat A$ possesses symmetry properties with respect to interachange of lead indices which are different from those of the frozen scattering matrix: \cite{MBmag05}
The symmetry of the $\hat A$ matrix depends on differences of the matrix elements of the $L$ and $R$
matrices [see Eqs. (\ref{Eqad_4C}) and (\ref{Eqad_4D})]. The main point is that the symmetry of the $\hat A$-matrix is fundamentally different from the (frozen) scattering matrix symmetry.

\section{Alternative unitarity conditions}
\label{appendixC}

We can obtain another (alternative) formulation of the unitarity condition for the matrix $\hat S_{in}$.
In the main part of the paper we use the resulting conditions to simplify the expressions for the pumped 
current.
Substituting Eq.\,(\ref{Eqdb_13A}) into Eq.\,(\ref{Eq_new2B}) and performing the inverse Fourier transformation we deduce (as before we treat $\hat{\cal S}^{(q,\tau)}$ as energy independent over the scale of order $\hbar\Omega$), 

\begin{equation}
 \label{m8_2}
\begin{array}{c}
\sum\limits_{q=0}^{\infty} \hat{\cal S}^{(q,\tau)}(t,E)\hat{\cal S}^{(q,\tau)\dagger}(t,E) \\
\ \\
+ \sum\limits_{p=0}^{\infty} \sum\limits_{s=1}^{\infty} e^{-2iskd} \hat{\cal S}^{(p,\tau)}(t,E) \hat{\cal S}^{(p+s,\tau)\dagger}(t + 2\tau s,E) \\
\ \\
+ \sum\limits_{q=0}^{\infty} \sum\limits_{s=1}^{\infty} e^{2iskd} \hat{\cal S}^{(q+s,\tau)}(t,E) \hat{\cal S}^{(q,\tau)\dagger}(t - 2s\tau,E)  = \hat I\,.
\end{array}
\end{equation}  
\ \\
These identities have to  hold for any wave number $k$. 
Taking into account that in our case the matrix $\hat {\cal S}^{(q,\tau)}$ can be kept energy independent over the interval corresponding to a change of $kd$ by $2\pi$, we derive next three equations from Eq.\,(\ref{m8_2}):

\begin{subequations}
 \label{m8_3} 
\begin{equation}
 \label{m8_3_A} 
\sum\limits_{q=0}^{\infty} \hat{\cal S}^{(q,\tau)}(t,E)\hat{\cal S}^{(q,\tau)\dagger}(t,E) = \hat I\,, 
\end{equation} 

\begin{equation}
 \label{m8_3_B} 
\sum\limits_{p=0}^{\infty} \hat{\cal S}^{(p,\tau)}(t,E) \hat{\cal S}^{(p+s,\tau)\dagger}(t + 2\tau s,E) = \hat 0\,,
\end{equation} 

\begin{equation}
 \label{m8_3_C} 
\sum\limits_{q=0}^{\infty} \hat{\cal S}^{(q+s,\tau)}(t,E) \hat{\cal S}^{(q,\tau)\dagger}(t - 2\tau s,E) = \hat 0\,,
\end{equation} 
\end{subequations}
\ \\
where $\hat 0$ is a null matrix. 
Notice that while the unitarity condition Eq.\,(\ref{Eqdb_15}) is quite general, the untitartity conditions obtained above are valid only within the approximations we made to get the matrix $\hat S_{in}$, Eq.\,(\ref{Eqdb_13A}). 

Taking a time derivative of Eq.\,(\ref{m8_3_A}) we immediately prove that Eq.\,(\ref{Eqdc_4B}) is real.
In addition using Eqs.\,(\ref{m8_3_B}) and (\ref{m8_3_C}) we can show that the product of scattering amplitudes ${\cal \hat S}^{(q,\tau_{0})}(t,\mu) {\cal\hat S}^{(q^\prime,\tau_{0})\dagger}(t-2\tau_{0}[q-q^\prime],\mu)$ (corresponding to particles leaving the scatterer at different time moments, $t$ and $t-2\tau_{0}[q-q']$) in fact does not contribute into an interference current $I_{\alpha}^{(i)}$, Eq.\,(\ref{Eqdc_4D}).


\begin{thebibliography}{11}
\bibitem{note1}
    See also Ref.\,\onlinecite{CMS07} where a close related decomposition into the temperature-independent (magnetic-flux-independent) and the temperature-dependent (magnetic-flux-dependent) parts for a current in the Mach-Zehnder interferometer was obtained.
\end{thebibliography}
\end{document}